\documentstyle[times,epsfig,chi2002b,graphics]{article}

\begin{document}

\title{SWiM: A Simple Window Mover}
\author{
        {\bf Tony Chang, Damon Cook, Ramona Su}\\
        {\small University of Illinois, Urbana-Champaign}\\
        {\small Digital Computing Lab}\\
        {\small 1304 W. Springfield Avenue}\\
        {\small Urbana, IL 61801}\\
        {\small $\lbrace$tychang1, djcook, ramonasu$\rbrace$@uiuc.edu}
}

\maketitle

\abstract
As computers become more ubiquitous, traditional two-\-di\-mensional interfaces must be replaced with interfaces based on a three-dimensional metaphor.  However, these interfaces must still be as simple and functional as their two-\-di\-men\-sion\-al predecessors.  This paper introduces SWiM, a new interface for moving application windows between various screens, such as wall displays, laptop monitors, and desktop displays, in a three-dimensional physical environment.  SWiM was designed based on the results of initial ``paper and pencil'' user tests of three possible interfaces.  The results of these tests led to a map-like interface where users select the destination display for their application from various icons.  If the destination is a mobile display it is not displayed on the map.  Instead users can select the screen's name from a list of all possible destination displays.  User testing of SWiM was conducted to discover whether it is easy to learn and use.  Users that were asked to use SWiM without any instructions found the interface as intuitive to use as users who were given a demonstration.   The results show that SWiM combines simplicity and functionality to create an interface that is easy to learn and easy to use.

\subsection{Keywords}
Ubiquitous computing, map, SWiM

\setcounter{secnumdepth}{5} 
\section{INTRODUCTION}
\label{s:intro}

As technology progresses, computers are extending beyond the desktop into the surrounding environment -- into walls, onto wrists, even into both large and small appliances \cite{ubicomp}.  As computers become more ubiquitous, new interface methods are required to interact with them.  For example, what methods of input work well for a wall monitor?  Do a mouse and keyboard still make sense in every context?  One key to success in integrating ubiquitous computing environments into everyday use is making interactions between users and technology seamless and fluid.  Therefore, new interface methods need to be easy to learn and use.

In this paper a small piece of this larger issue is examined.  Imagine a room that contains multiple wall mounted displays, hand-held devices, and laptop computers.  Applications are no longer confined to one of these displays but can be moved freely from one to the next.  The traditional method for moving a window, selecting and dragging the title bar, is no longer viable because the displays are not contiguous.  SWiM, a \textbf{S}imple \textbf{Wi}ndow \textbf{M}over, is a first attempt at finding an intuitive way to replace this method in a multiple display ubiquitous computing environment.

One motivation for creating such an interface is to facilitate discussion among people using the room both in informal and formal settings.  For example, members of a group could copy an application window to multiple screens and work with the application simultaneously.  Furthermore, during a group brainstorming session subgroups could each work with multiple applications on separate displays and then bring all the applications back to a central display for a joint discussion.  Likewise, someone giving a presentation can eliminate the need to toggle between various applications by using multiple displays.  This presenter could also move windows that are less important to displays further away from the audience while moving more important windows closer.  

An intuitive interface is not achieved through the use of a specific algorithm or procedure.  Instead, it must be designed from a user's perspective.  After narrowing down from three possible approaches with early user testing, SWiM, a prototype interface, was implemented.  The design of SWiM incorporates ideas from the early user testing of these conceptual interfaces.

SWiM was designed and tested on top of the Gaia infrastructure in the ``Active Spaces Lab'', a ubiquitous computing laboratory.\cite{gaia}  This room contains plasma and projection displays on multiple walls, as well as laptop and desktop computers in various parts of the room.

In section 2 of this paper, we discuss work related to SWiM.  Section 3 relates the contributions of SWiM while section 4 discusses its implementation.  Section 5 shows the results of user tests of the interface and section 6 mentions some of the more pertinent lessons learned.  Finally, section 7 revels our conclusions about SWiM and section 8 tells what future work there is to do.

\section{RELATED WORK}
\label{s:related}
Moving application windows among various displays has\linebreak been the focus of research in multiple ubiquitous computing environments.  In i-Land, a room with an interactive electronic wall (DynaWall), computer-enhanced chairs, and an interactive table, three methods were introduced for moving application windows on the DynaWall.\cite{iland,dynawall}  Two of these methods, shuffling and throwing, are implemented using gestures.  Shuffling is done by drawing a quick left or right stroke above the title bar of a window.  This will move the window a distance equal to the width of the window in the gestured direction.  Throwing is done by making a short gesture backward, then a longer gesture forward.  This will move the window a distance proportional to the ratio between the backward and forward movement.  The throwing action requires practice because there is no clear indication of how far something will move prior to using it.  The final method for moving windows in i-Land is taking.  If a user's hand is placed on a window for approximately half a second, that window shrinks into the size of an icon.  The next time the user touches any display, the window will grow behind the hand back to its original size.

In Stanford's iRoom, the PointRight system allows users to use a single mouse and keyboard to control multiple dis\-plays.\cite{pointright}  Changing displays is accomplished by simply moving the cursor off the edge of a screen.  Currently, iRoom does not move applications across displays, but this mouse technique could be extended to dragging application windows as well.

Another approach for manipulating objects (text, icons and files) on a digital whiteboard is ``Pick-and-Drop''.\cite{pickanddrop}  Using Pick-and-Drop, the user can move an object by selecting it on a screen with a stylus (a small animation is provided where the object is lifted and a shadow of the object appears) then placing it on another screen by touching the desired screen with the again.  The benefits of this approach include a more tangible copy/paste buffer and a more direct approach than using FTP or other file transfer techniques.

A more general method for controlling devices in a ubiquitous computing environment is Microsoft's XWand.\cite{xwand,xwand2}  The XWand is a wireless sensor package in the shape of a wand that senses its own orientation with respect to the room.  While the original idea is to use it in combination with gestures to control various devices (stereo, TV, lights, etc), it could also be extended to move application windows between multiple displays.  This new concept could then be incorporated with the ideas in PointRight by allowing various input devices to be used, such as the XWand.

\section{CONTRIBUTIONS}
\label{s:contrib}

\begin{figure}
	\begin{center}
		\epsfbox{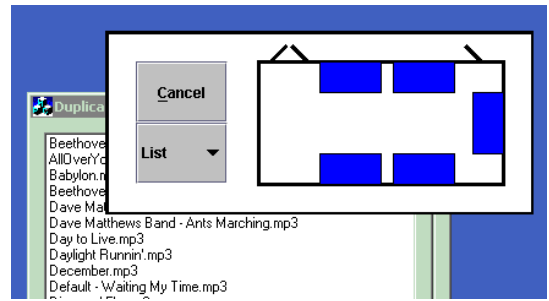}
	\end{center}
	\caption{Pressing an extra button on the titlebar activates SWiM.  SWiM provides a map of the room and a list of all possible destination displays.  Choosing an icon or a member of the list moves the applications to the appropriate screen.}
	\label{f:swim}
\end{figure}
To aid in moving windows between various displays in a three-dimensional ubiquitous computing environment, we\linebreak have developed SWiM.  Rather than using a menu to select the window to be moved, an additional button is added to the title bar of each application.  This button is used to present the user with a map of the room (Figure \ref{f:swim}).  On the map, each of the static displays in the room are represented by blue rectangles.  Other landmarks, such as doors, are also on the map to help the user orient the user in the room.  The user can then select a destination display on the map by clicking on it to move the application window.

For dynamic room elements, such as PDAs or laptop computers, a list is provided that shows all displays in the room including those on the graphical display.  Selecting an item from the list will also move the application to the appropriate display.

\section{METHOD}
\label{s:method}
To create SWiM, three design concepts were presented to six users in early user testing.  The comments provided by the users were then incorporated into the design and implementation of SWiM.  After SWiM was implemented, a usability study was performed on the interface.  Section \ref{method:early} describes the designs of the interfaces tested in early user testing.  Section \ref{method:testing} describes the methodology of the early user testing.  Section \ref{method:results} describes the results from the early user testing.  Finally, section \ref{method:dev} describes the development of SWiM based off the results found in Section \ref{method:results}.   

\subsection{Early User Design}
\label{method:early}
\begin{figure*}
	\begin{center}
\includegraphics[width=0.9\textwidth]{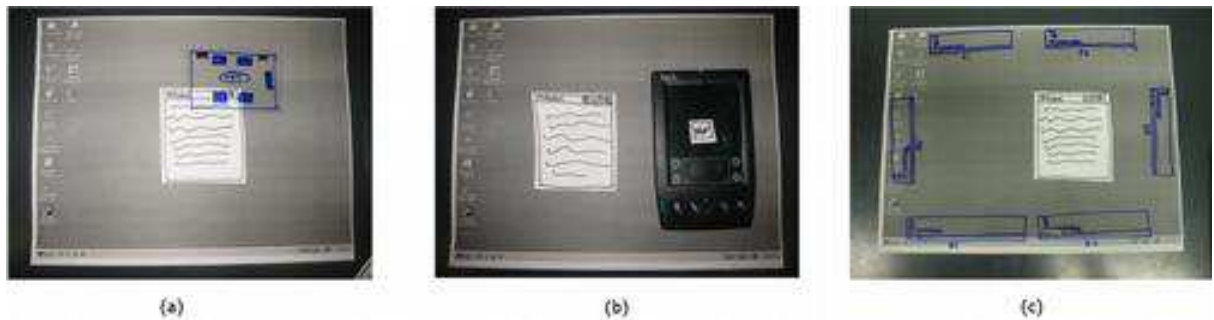}
	\end{center}
	\caption{The three initial designs used to develop SWiM: (a) the pie menu interface, (b) the artist palette interface, (c) the throw interface.}
	\label{f:paperpencil}
\end{figure*}

The initial design stage consisted of making ``paper and pencil'' designs of three possible interfaces for intuitive movement of applications between screens in a ubiquitous computer environment.  The three interfaces that were used in the early user testing were the pie menu, artist palette, and the throw interface (Figure \ref{f:paperpencil}).

Figure \ref{f:paperpencil}a shows the pie menu interface.  This interface is activated whenever a user drags an application window on the screen.  When this happens, a menu pops up to surround the cursor.  The menu uses icons to represent various screens available in the room.  Additionally, a list icon is present to show all available destinations by name.  Other reference points, such as doors or windows, are also displayed around the menu to help orient the user.  To move the window to another screen, the user simply releases the mouse over the appropriate icon.  To choose not to move to another screen, the user has only to move outside the range of the menu.

Figure \ref{f:paperpencil}b shows the artist palette interface.  This interface is activated by pressing an extra button on the title bar of the application window.  Activation causes ``DROP'' buttons to appear on all the displays in the room.  Pressing the ``DROP'' button on a display causes the application window to move to that display.  If the user presses the ``DROP'' button on a PDA, the PDA will save the application as a small icon on the PDA.  The user can then click a ``SEND TO'' button on the PDA, which will list all the available displays in a menu.  The user then selects the desired display.

The throw interface (shown in figure \ref{f:paperpencil}c) allows users to move application windows by ``throwing'' them at the desired display (clicking and holding on the title bar, moving in the appropriate direction and then releasing the mouse button).  To activate this interface, a user must click on the desired window and hold for a brief time.  When the interface is activated, rectangular icons appear along the edges of the desktop.  If the user is on a laptop, the up direction indicates that they want to move the window to a display on the walls of the room.  If the user is on a wall display, the up direction indicates displays on the other side of the room.  Throwing to the images on the left and right represent moving the application to the display to the left or right of the current display.  Throwing down is only used when the user is on a wall display, and is trying to move the application onto a display not on the wall such as a laptop on the table in the center of the room.

\subsection{Method of Testing}
\label{method:testing}
To test the three interfaces described in section \ref{method:early}, six users evaluated each interface by performing three tasks using ``paper and pencil'' prototypes.  The three tasks were moving an application from laptop to a wall display, a wall display to another wall display, and laptop to a wall display.  The three interfaces were presented in a different order for each user to prevent learning effect.  To provide consistency between each user test, each user was given an introduction to the lab as well as an introduction to the project.  Before each interface, the name of the interface was given and the user was asked to verbalize how they would complete each task.  The evaluation of each interface was based on the comments a user gave while completing the task for a given interface.  Additional information was provided through a user's overall experience with an interface.

\subsection{Results of Early User Testing}
\label{method:results}
With the first interface tested (whether the pie, throw, or\linebreak artist), five of the six users would try to drag the application off the screen in the direction of the display they wished to move to.  Then, after completing the tasks with the initial interface, the users tended to use their experiences from the previous interfaces to move the application window in the new interface.  

With the pie interface, three of the six users had trouble associating the menu of the displays with the physical displays in the room.  The other users were able to associate the rectangles of the menu with the displays on the walls of the room.  Even with this recognition, the users were still confused and disoriented with which display corresponded with the map items.  Another problem users faced with this interface was their confusion with the displays that were not located on the map.  When the users were asked to move the window back to the laptop from a plasma screen, two users had problems using LIST in the interface because they assumed that since the laptop was not shown in the map that they were not able to move the screen back onto the laptop.  

Some things that users liked about the pie interface were that the interface was intuitive and the way the menu appeared.  Some things that users did not like about this interface were that it was hard for them to figure out what the ``blue screens'' were.  They also had difficulties understanding the figures used as landmarks.  Three users thought that you either could do things with the landmarks or did not understand what the landmarks represented.  Users did not like the list feature because it was not immediately obvious that items not located in the map were located in the list.  Others thought that the list was redundant.  Furthermore, users were concerned that if  they were in a room that was symmetric, the orientation of the pie menu would be unclear.  

The users made several suggestions on how this interface might be improved.  The users suggested that instead of having the menu pop up when the application is moved (which can prove to be annoying), it could have another button on the title bar (next to the minimize, maximize, and close buttons) to indicate the explicit action of moving the window.  This could help provide the user with a visual cue.  To make an easier association for the displays in the menu with the physical displays, users suggested labeling both the physical and menu displays clearly.  This could help alleviate problems in a symmetrical view.

In the artist palette interface, four of the users were aware of the extra button and associated the action of the button with an indication to move the application window.  Four users tended to hit the drop button immediately on the display where they wished to drop their window.  When it was explained how the PDA could be used instead, most users liked the idea of minimizing the movement that they would have to make.  

Users liked having an extra button to cause an explicit action on this interface.  They also liked how movement was minimized.  Some things that users did not like about this interface were that ``DROP'' and ``SEND TO'' were not descriptive enough to communicate the purpose of the buttons.  Other shortcomings of this interface noted by the users include the extra movement needed if a user did not have a PDA, moving an application became a sequence of steps rather than one flowing step, and moving a lot of windows could be burdensome.  

Some suggestions that were made for this interface were to have the extra button point in different directions depending on the display that a user was currently at: up for moving onto the plasma and down for moving onto the laptop.  Also, it was suggested that displays should have more detailed labels.

In the throw interface, five of the users had difficulty identifying what action would start the process of moving an application window.  It was not clear to the users that clicking and holding the title bar would bring up the menu.  After the users discovered this, they had difficulty in understanding the ``throw'' aspect of the interface (holding the title bar, moving it towards the display menu item, and then letting go).  Another problem with this interface is that most of the users did not associate the location of the menu items with the physical locations of the displays.  The users generally looked at the display name in the menu, and moved the window towards that screen.  

Most users like the labeling of the menu items (the rectangular displays located at the edges of the screen), and thought the interface was more ``intuitive'' than the others.  The users disliked the action of clicking and holding on the title bar to activate the interface.  In addition, users found that dragging the application all the way to a menu item was more desirable than throwing the window towards it. 

The users also posed general questions about this interface that aided in the design of SWiM.  These questions touched on ideas such as what would happen if a user just wanted to move the display window around and where the window would pop up on the destination display.

\subsection{Development}
\label{method:dev}
Based on the results of the early user tests, SWiM was implemented in a similar way to the pie menu.  To reduce some of the shortcomings found during the initial user test, an extra button was added to the title bar to make movement of the window more explicit. To help eliminate problems with canceling the action, a ``Cancel'' button was added to the interface.  Additionally, the menu was split into two parts: one for the buttons (Cancel and List) and one for the map of the room.  The map portion is designed so that it resembles the room layout.  A rectangle surrounds the map of the room to symbolize its borders.  To show doors as landmarks, the menu uses lines, similar to those used by architects to symbolize doors on floor plans.  The menu is implemented in Java, and the movement of windows uses a call to a library in the Gaia infrastructure, which already implements the movement of windows between screens.

\subsection{Evaluation}
\label{method:eval}
After implementing the interface, a usability study was conducted on the SWiM interface.  Eight users participated in the test and they were divided into two groups-- trained and untrained.  In the untrained group, each user was given a very high level, general overview of the interface.  They were told that there was an extra button on the title bar, and after the button was pressed the interface would appear, allowing them to move the application to another screen.  With the trained group, detailed instructions of how to use the map and the list portions of the interface was given.  Along with verbal instructions, the users were also provided a demonstration of how to use the interface.  In both groups, the users were asked to perform four tasks.  The four tasks were:

\begin{enumerate}
	\item to move an application window from a given wall display, to another wall display across the room
	\item to move an application window from a given wall display to a given display on the table
	\item to move an application window from a display on the table in the center of the room, to a given wall display
	\item to move an application window from one display on the table to the another display on the table
\end{enumerate}
The tasks were ordered in a Latin square for each group of users (trained and untrained).  Before starting the tasks, the users were allowed to familiarize themselves with the touch-screen interface of the plasma displays on the walls and the stylus of the WACOM digitizing displays on the tables.  The users were also instructed that during the test they should first try to move the windows to the desired displays before asking the test giver for help.  After the users completed each task, they were asked to fill out a questionnaire.  Each questionnaire asked a user what they expected to result from a given action and describe how to complete a desired task.  The questionnaire also requested comments about what the user liked, disliked, how the user would change the interface, and a ranking of their perception of the difficulty of each task.

\section{RESULTS}
\label{s:results}
When the users were asked questions about how to operate the interface or what the results of specific actions would be in the interface, they were able to answer correctly.  The only exception was one response that indicated the wrong destination screen when a specific icon was selected.  From the limited interaction that each user had with the interface, the users seemed to understand it well.  This indicates that the simple interface of SWiM is easy to learn.

\begin{figure}
	\begin{center}
		\epsfbox{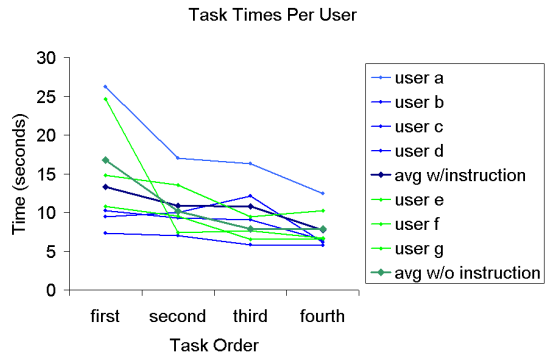}
	\end{center}
	\caption{This graph shows the time taken by each user to complete the four given tasks in order.  The tasks were randomly assigned to the users in such a way that each user did a different first task than the other users in their category.  The thick lines represent the average times for each category.}
	\label{f:tasktime}
\end{figure}

When asked to evaluate what was the hardest to do, most users said that moving something from a wall display (to a wall or desktop) or moving something to a desktop display (from a wall or desktop) was the most difficult.  Based on user comments, moving something from a wall display to another wall display was difficult because the users needed to reorient themselves when looking at the map.  Moving from a wall display or a desktop display to a desktop displays seemed difficult for users because the desktop displays were not located on the map of the room.

Figure \ref{f:tasktime} shows the time taken for users to complete tasks in the order they were received.  There is a general decrease in time taken by users as they accomplish more tasks.  The difference between the average times were within two seconds for this small sample set.  This decrease in time as a user interacts with SWiM indicates that there is a short learning curve to this interface.  With a short learning curve, SWiM promotes interaction that is seamless and fluid.  Once the user is familiar with the interface, moving a window does not seem to occupy a user's attention.

\begin{figure}
	\begin{center}
		\epsfbox{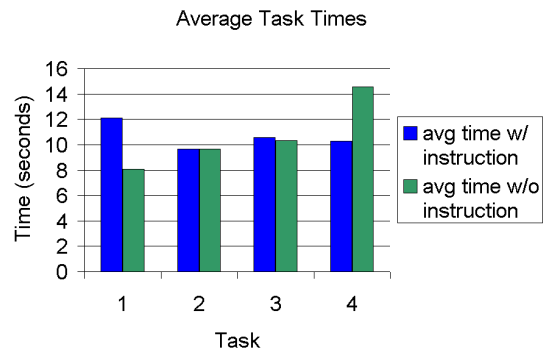}
	\end{center}
	\caption{This graph shows the time taken on average for specific tasks during the user tests.  Task 1 was to move between wall displays on opposite sides of the room, task 2 was to move from a wall display to a desktop display, task 3 was to move from a desktop display to a wall display, and task 4 was to move from one desktop display to another.}
	\label{f:avgtime}
\end{figure}

Figure \ref{f:avgtime} shows the average times taken by each group of users for each specific task.  In fact, the first task took the instructed group longer on average while the fourth task took the uninstructed group longer.  There was no statistical difference in times between the untrained user group and the trained user group.  The conclusion that can be drawn from this is that SWiM is an interface that is simple to use.  It is an interface that a user can learn with a few interactions with the interface.  The user test also supports this idea in that no user required assistance to move a window from the test giver during the experiment.    

Although there were four users in each category of the testing, one user's data was excluded due to technical problems during the experiment.

\section{LESSONS}
\label{s:lessons}
User comments revealed several important lessons.
\begin{itemize}
	\item \textit{It is difficult for users to orient themselves with regard to a map.}  The map that was used in SWiM used the up direction to mean north.  Almost all users felt that up should actually indicate the direction that they were facing.  Additionally, several users noted that landmark indicators were valuable in orienting themselves.  However, it was observed in ``paper and pencil'' testing that unclear indicators were generally ignored.  Along the same lines, many users indicated that a clear ``You Are Here'' icon is important.
	\item \textit{As much information as possible should be presented to the user, but not at the expense of readability.}  In the experiment, desktop displays were not graphically represented in the map.  A majority of users commented that those displays should have been visible in some way.  Although this might not take the form of a specific icon for every possible display, some indication that there are available displays is necessary.
	\item \textit{Users like flexibility in the interface.}  In SWiM, users have the option of using the list or map portion of the interface.  Allowing such flexibility allows the interface to conform to various users' needs.  For instance, in a group environment, it might be easier for a user to move screens using a list when another group member requests a window to be on a particular screen.  On the other hand, for a presenter, a map would be a better choice because they would be able to orient an application's current location more quickly than reading a display's name and highlighting in on a list.  
	\item \textit{There should be an easy way to undo whatever action has just been taken.}  Although the users did not make mistakes that resulted in a need to undo their action, it was clear from pilot testing that errors in this system have an exceptionally high cost to correct.  Specifically, having to walk across a room to try an action again negates the quickness and ease of the system.
	\item \textit{Other functionality might be added to the interface.}  SWiM is currently limited to moving only one application window.  Allowing the capability of duplicating a window to another display can be very beneficial for a user.  
	\item \textit{Touch screens can be difficult to use and should be compensated for.}  Several users in this study had difficulty in pressing the correct position on the wall displays that were used.  Additional complications came from users that were unfamiliar with the environment.  Several users had to touch the screen multiple times before finally activating the desired location.
\end{itemize}
\section{CONCLUSION}
\label{s:conclusion}

Some of the outside factors that may have caused variance in the data include the height of the subjects (some shorter people had difficulty using the wall displays), level of familiarity with the environment, and the small number of users in each group.

The data shows that users understood the system comparatively, whether they were instructed in its use or not.  Joining this with specific comments from the user tests, this shows that users are able to quickly grasp the interface because it is based on the physical metaphor of a map.

SWiM is an interface that has been successful in moving application windows in a three-dimensional environment.

\section{FUTURE WORK}
The response from the testing of SWiM indicates that while it is essential to base the interface on a common metaphor, it would also be useful to incorporate more information from the environment into that metaphor.  The next iteration of the interface would show more information about what is on a screen on its icon, as well as more color and landmarks to help orient the user.

While the focus of this experiment was to move applications from the current display, further research will explore ways to move applications on any display.  The lessons learned from SWiM can be extended to a map that show icons for each application on every available display.  Moving the window can then be done by dragging a window icon to the desired display icon.

Finding other methods of representing a three-dimensional environment can also be explored.  The limitations of the current map is that displays cannot easily be stacked vertically in the space.  The map may also require a large portion of the screen on small displays, such as a PDA.  Finally, working to allow the map to be dynamically generated to handle mobile computers is necessary.

\bibliographystyle{alpha}

\begin{thebibliography}{00}

\bibitem{dynawall}
Geibler, J.
\newblock Shuffle, throw or take it! Working efficiently with an interactive wall, \textit{CHI'98 Summary}, pp.265-266.

\bibitem{usable}
Gould, J.D.
\newblock How to Design Usable Systems, in \textit{Readings in Human-Computer Interaction: Toward the Year 2000}, pp. 93-121, 1995. 

\bibitem{pointright}
Johanson, B., G. Hutchins, T. Winograd, and M. Stone.
\newblock PointRight: Experience with Flexible Input Redirection in Interactive Workspaces, in \textit{Proceedings of ACM Conference on User Interface and Software Technology}, 2002, pp. 227 - 234.

\bibitem{xwand}
Krumm, John, Steve Shafer, and Andy Wilson.
\newblock  How a Smart Environment Can Use Perception, unpublished, Presented at Ubicomp 2001, Workshop on Sensing and Perception for Ubiquitous Computing, September 2001.

\bibitem{usable2}
Mack, R.L and J. Nielsen.
\newblock Usability Inspection Methods: Executive Summary, in \textit{Readings in Human-Computer Interaction: Toward the Year 2000}, pp. 170-181, 1995.

\bibitem{pickanddrop}
Rekimoto, Jun.
\newblock Pick-and-Drop: A Direct Manipulation Technique for Multiple Computer Environments, in \textit{Proceedings of UIST'97}, pp. 31-39, 1997

\bibitem{}
Roman, Manuel, and Roy H. Campbell.
\newblock A User-Centric, Resource-Aware, Context-Sensitive, Multi-Device Application Framework for Ubiquitous Computing Environments,  \textit{Technical Report UIUCDCS-R-2002-2284 UILU-ENG-2002-1728, University of Illinois at Urbana-Champaign}, July 2002.

\bibitem{}
Roman, Manuel, et al.
\newblock Gaia: A Middleware Infrastructure to Enable Active Spaces, in \textit{IEEE Pervasive Computing}, Oct-Dec 2002:  74-83.   

\bibitem{gaia}
Roman, Manuel, et al.
\newblock GaiaOS: An Infrastructure for Active Spaces, \textit{Technical Report UIUCDCS-R-2001-2224 UILU-ENG-2001-1731}, University of Illinois at Urbana-Champaign, May 2001.

\bibitem{iland}
Streitz, Norbert A., et al.
\newblock i-LAND: an interactive landscape for creativity and innovation, in \textit{Proceedings of the SIGCHI conference on Human factors in computing systems}, 1999, pp. 120-127.

\bibitem{ubicomp}
Weiser, Mark. 
\newblock Hot Topics:  Ubiquitous Computing, \textit{IEEE Computer}, October 1993.

\bibitem{xwand2}
Wilson, Andrew and Steven Shafer.  
\newblock  XWand: UI for intelligent spaces, in \textit{Proceedings of the SIGCHI conference on Human factors in computing systems}, 2003, pp. 545-552.

\end{thebibliography}

\end{document}